\documentclass[prl,showpacs,amsmath,amssymb,twocolumn]{revtex4-2}

\usepackage{graphicx}
\usepackage{dcolumn}
\usepackage{bm}
\usepackage{epstopdf}
\usepackage{upgreek}
\usepackage{amsmath,amssymb,amsthm,commath,esint,tikz-cd,tikz, mathtools}
\usepackage{array}

\numberwithin{definition}{section}

\numberwithin{proposition}{section}

\numberwithin{theorem}{section}

\numberwithin{lemma}{section}

\numberwithin{corollary}{section}
\theoremstyle{remark}

\numberwithin{example}{section}

\usepackage{hyperref}
\hypersetup{colorlinks=true, citecolor=blue, urlcolor=blue, linkcolor=blue}

\usepackage{comment}

\begin{document}

\title{Real-space superconducting properties in the atomically-thin limit: \textit{Ab initio} approach and its application to Josephson junctions}

\author{Jonas Bekaert}
\email{jonas.bekaert@uantwerpen.be}
\affiliation{
 Department of Physics, University of Antwerp,
 Groenenborgerlaan 171, B-2020 Antwerp, Belgium
}

\author{Mikhail Petrov}
\thanks{Present address:  Department of Mechanical Engineering, Tufts University, Medford, MA 02155, USA}
\affiliation{
 Department of Physics, University of Antwerp,
 Groenenborgerlaan 171, B-2020 Antwerp, Belgium
}

\author{Milorad V. Milo\v{s}evi\'c}
\email{milorad.milosevic@uantwerpen.be}
\affiliation{
 Department of Physics, University of Antwerp,
 Groenenborgerlaan 171, B-2020 Antwerp, Belgium
}


\begin{abstract}
\noindent  
Real-space superconducting properties are increasingly important to characterize low-dimensional, layered, and nanostructured materials. Here, we present a method to extract the real-space superconducting order parameter from the superconducting gap spectrum obtained via anisotropic Migdal-Eliashberg calculations, using the Bloch wave functions of the Fermi states. We apply this approach to a selection of atomically thin material systems. Our analysis of gallenene, a monolayer of gallium atoms, shows that its planar and buckled phases exhibit distinct superconducting order parameter behaviors, shaped by their structural and electronic properties. Furthermore, we demonstrate that our real-space approach is exceptionally suited to identify and characterize Josephson junctions made from van der Waals materials. Our examination of a bilayer of NbSe$_2$ reveals that the van der Waals gap acts as an intrinsic weak link between the superconducting NbSe$_2$ layers. Therefore, a bilayer of NbSe$_2$ represents one of the thinnest and most tunable Josephson junction architectures, with potential applications in quantum devices. Our findings underscore the utility of transformation into real-space in understanding superconducting properties through \textit{ab initio} calculations. 
\end{abstract}

\date{\today}
\maketitle

Recent years have seen a surge in research on superconductivity in low-dimensional systems, such as two-dimensional (2D) materials. A defining feature of these low-dimensional superconductors is their extraordinary tunability, which significantly exceeds that of bulk materials. This tunability can be achieved \textit{(i)} chemically, via surface decoration \cite{PhysRevLett.123.077001,PhysRevB.105.245420,D2NR01939F,HAN2023100954,Soskic2024,Seeyangnok_2024}, \textit{(ii)} structurally, through the application of strain \cite{PhysRevB.96.094510,PhysRevLett.123.077001,Soskic2024,Chen2020,doi:10.1021/acs.nanolett.4c02421}, or \textit{(iii)} electrically, through applied gating \cite{deVries2021,Rodan-Legrain2021,Marini_2023,Soskic2024}. Alongside these efforts to optimize individual monolayer materials, extensive research is focused on exploring the benefits of integrating van der Waals (vdW) monolayers into advanced multilayers and heterostructures, to tailor superconducting properties with precision and control. For example, the 6R-phase of the transition metal dichalcogenide (TMD) tantalum disulfide (TaS$_2$) features alternating 1T and 1H monolayers, which are polymorphs with a different sulfur coordination. The insulating 1T layers, governed by a charge density wave state, confine 2D superconductivity to the 1H planes \cite{Achari2022,PhysRevLett.133.056001}. Another TMD-based vertical superlattice features alternating 1H-TaS$_2$ and Sr$_3$TaS$_5$ layers, giving rise to an anisotropic fermiology and incommensurately modulated striped electronic states \cite{Devarakonda2024}.

To better understand the anisotropic and spatially modulated electronic and superconducting properties of these emerging 2D materials and heterostructures, the prevailing atomistic theoretical models do not suffice, as they fail to directly connect to real-space properties. The most widely used \textit{ab initio} approach to studying phonon-mediated superconductivity, based on Migdal-Eliashberg theory, determines the superconducting gap spectrum $\Delta_{n\mathbf{k}}$ for electronic bands at the Fermi surface (band index $n$), expressed in momentum space (wave vector $\mathbf{k}$) \cite{Eliashberg1,Eliashberg2,RevModPhys.89.015003}. As such, the \textit{ab initio} Migdal-Eliashberg framework has largely ignored the real-space behavior and spatial extent of the superconducting order parameter. In 2015, an alternative \textit{ab initio} approach, density functional theory for superconductors (SCDFT), was used to compute real-space superconducting order parameters of systems of interest at that time, such as MgB$_2$, Ca-intercalated graphite and hole-doped graphane \cite{PhysRevLett.115.097002}. More recently, computational methods were developed that incorporate real-space effects using the Bogoliubov-de Gennes framework, however, these are based on semi-empirical parameters rather than the true superconducting pairing interactions \cite{PhysRevB.105.125143,PhysRevB.110.134505}. 

As a result, the influence of real-space effects on phonon-mediated superconducting materials, particularly under dimensionality reduction, remains a largely unexplored frontier within atomistic \textit{ab initio} calculations. In particular, this limitation hinders the theoretical characterization of systems where the spatial distribution of superconducting order is a key property, such as Josephson junctions composed of vdW materials, which are projected to further advance superconducting quantum computing devices \cite{Liu2019}. 

Therefore, in this Letter, we introduce a computational framework that leverages widely available Migdal-Eliashberg results -- such as those provided by the Electron-Phonon Wannier (EPW) code \cite{Giannozzi_2017,PhysRevB.88.085117,PONCE2016116} -- to efficiently evaluate real-space superconducting properties. We apply this approach to two low-dimensional superconducting systems: two gallenene phases in monolayer form and a homobilayer of the transition metal dichalcogenide (TMD) NbSe$_2$. Our results highlight key advantages of a real-space perspective, including its ability to link chemical bonding to superconducting properties, characterize the evanescence of the superconducting state in the out-of-plane direction of 2D materials, and describe the interlayer distribution of the superconducting state in multilayer structures -- an aspect particularly relevant to the Josephson effect and its applications.

The real-space superconducting density (or order parameter) $\chi(\mathbf{r},\mathbf{r}')$ is a two-particle function that depends on two spatial coordinates, as a direct result of the nature of the Cooper pair. It can be conveniently described as a function of the center-of-mass coordinate $\mathbf{R}=(\mathbf{r}+\mathbf{r}')/2$ and relative coordinate $\mathbf{s}=\mathbf{r}-\mathbf{r}'$ \cite{PhysRevLett.115.097002}. The $\mathbf{R}$-dependence corresponds to the distribution of the Cooper pairs, while the dependence on $\mathbf{s}$ reflects the internal structure of the pairs. Hence, the Cooper-pair distribution in real space can be calculated as follows \cite{PhysRevLett.115.097002}, 
\begin{equation}
\chi(\mathbf{R},\mathbf{s}=0)=\sum_{n\mathbf{k}} \frac{\Delta_{n\mathbf{k}}}{2 E_{n\mathbf{k}}} \tanh\left(\frac{\beta E_{n\mathbf{k}}}{2}\right) \left|u_{n\mathbf{k}}\left(\mathbf{R}\right)\right|^2 ~. \\
\label{Eq:chi}
\end{equation}
Here, $E_{n\mathbf{k}}=\sqrt{\xi_{n\mathbf{k}}^2+\Delta_{n\mathbf{k}}^2}$ represents the Bogoliubov spectrum, where $\xi_{n\mathbf{k}}$ is the normal-state electronic dispersion, and $u_{n\mathbf{k}}$ denotes the lattice-periodic Bloch wave functions, making $\chi(\mathbf{R},\mathbf{s}=0)$ inherently lattice-periodic. In contrast, the $\mathbf{s}$-dependence is not lattice-periodic and decays over the characteristic pair coherence length \cite{PhysRevLett.115.097002}. Our computational framework to efficiently obtain and combine all the quantities in this expression, entitled `Code for Real-space SuperConductivity' (CRESCO), is discussed in the Supplementary Material (SM). 

\begin{figure}[t]
                \includegraphics[width=\linewidth]{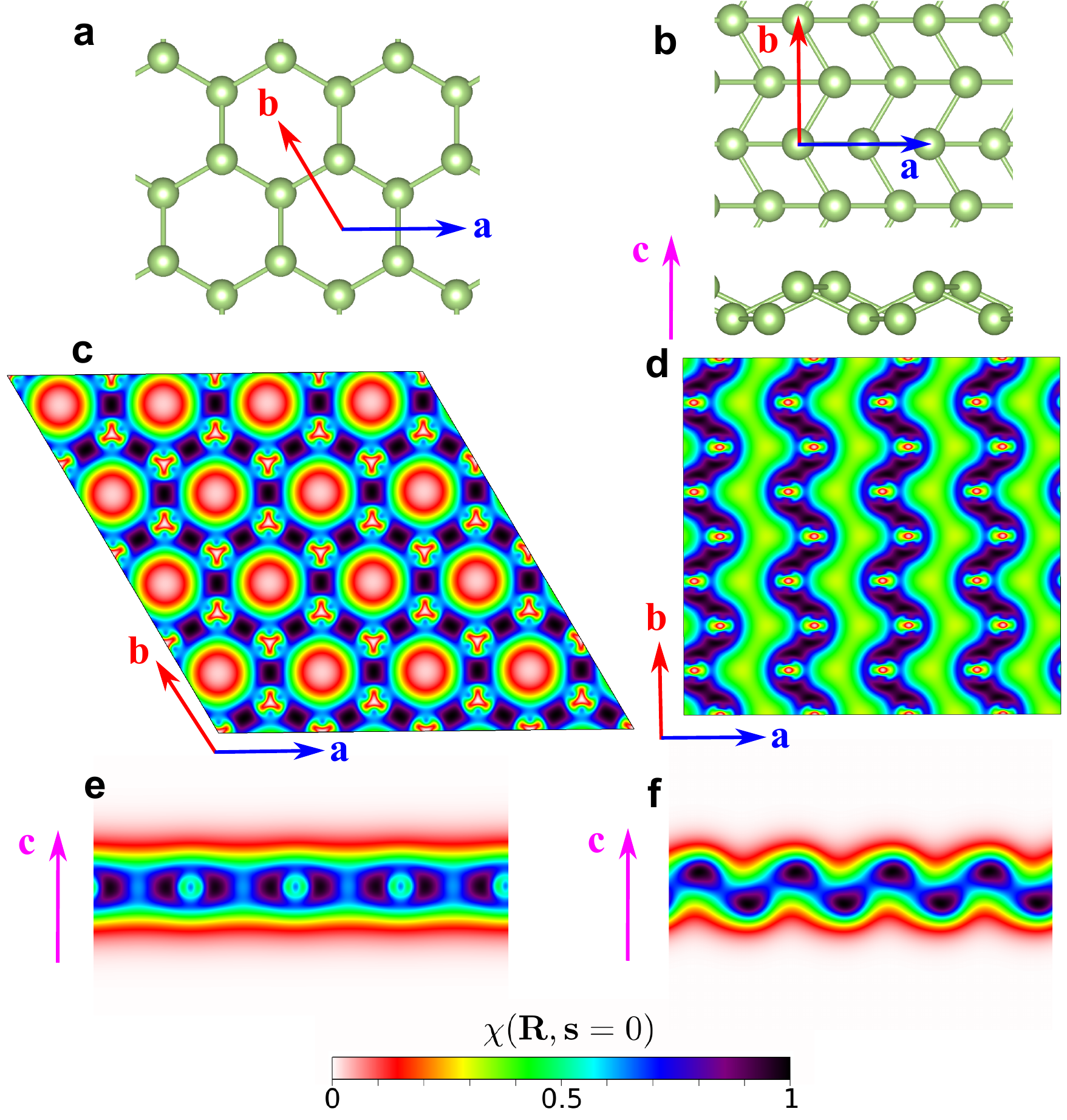}
                \caption{Crystal structures of two 2D gallium structures (called gallenene) \cite{Petrov_2021}: (a) Ga-100 and (b) Ga-010 gallenene. Normalized real-space superconducting density $\chi(\textbf{R},\mathbf{s}=0)$ for the in-plane (c,d) and out-of-plane (e,f) directions of Ga-100 and Ga-010, respectively.}
                \label{fig:gallenene}
\end{figure}

As a first example, we consider a monolayer composed of gallium atoms, called gallenene. Gallenene has two polymorphs, Ga-100 and Ga-010 (Figure \ref{fig:gallenene}a–b), which can be transferred on various substrates \cite{Kochate1701373}. Our earlier \textit{ab initio} calculations have revealed intrinsic phonon-mediated superconductivity in both gallenenes, with critical temperatures in the range of 7--10 K -- exceeding that of bulk $\alpha$-gallium \cite{Petrov_2021}. Ga-100 exhibits a three-gap superconducting state, whereas Ga-010 features a single anisotropic superconducting gap \cite{Petrov_2021}.

By combining these calculated superconducting gaps with the matching Bloch wave functions according to Equation \eqref{Eq:chi}, we have obtained the Cooper-pair distributions, shown in Figure \ref{fig:gallenene}, in normalized form. Note that Equation \eqref{Eq:chi} can provide the distribution at arbitrary temperatures; however, we focus here on the $T \rightarrow 0$ limit, as it represents the fully developed superconducting state. Hence, the distributions were obtained at temperatures well below the compound's $T_c$ where the gaps are fully converged: 1 K for Ga-100 and 3.5 K for Ga-010, corresponding to their respective $T_c$ values of 7 K and 10 K. 

The in-plane Cooper-pair density in Ga-100, shown in Figure \ref{fig:gallenene}c, clearly reflects the $sp^2$ hybridization network, giving rise to the honeycomb crystal structure of gallenene. The highest Cooper-pair density is observed at the centers of the Ga-Ga bonds, while the density gradually diminishes toward the centers of the honeycomb facets. In contrast, the corrugated structure and zigzag arrangement of atoms in Ga-010 result in a markedly more anisotropic in-plane Cooper-pair distribution, as shown in Figure \ref{fig:gallenene}d. Here, the in-plane density corresponds to a cross-section of one of the two Ga layers. A slice through the other layer would yield an equivalent distribution, albeit with the regions of minima and maxima interchanged.

As discussed in the introduction, extensive research efforts are ongoing to develop advanced heterostructures integrating 2D superconducting materials. Hence, it is important to characterize the extent and evanescence of the Cooper pairs in the out-of-plane direction, which lies at the basis of superconducting interlayer hybridization and Cooper-pair-tunneling. To quantify this, we define the superconducting layer thickness $d$ as the distance between the isosurfaces where the Cooper-pair density has dropped to 10\% of the maximum $\chi$ value (normalized to 1 in this work). These correspond to the red regions on either side of the Ga planes in Figure \ref{fig:gallenene}e,f. 

For the planar polymorph Ga-100, the superconducting layer thickness is $d_{100} = 3.4$ \AA. In contrast, the isosurfaces of Ga-010 exhibit a wave-like shape due to its corrugated structure. Despite this, the superconducting layer thickness remains relatively uniform across the layer, with an average value of $d_{010} = 4.1$ \AA. Notably, the interlayer distance between the two Ga planes in the Ga-010 structure is 1.2 \AA, which only partially translates into a larger $d$ value compared to the planar Ga-100. The superconducting layer thickness is an important property to predict proximity effects on monolayer materials interfaced with a substrate material \cite{Kim2017,Dreher2021,PhysRevB.109.L140503}. Altogether, these findings highlight the strong spatial variations in the Cooper-pair distribution in 2D materials and underscore the crucial role of the superconducting gap function and Bloch wave functions in shaping this behavior. 

\begin{figure}[t]
                \includegraphics[width=\linewidth]{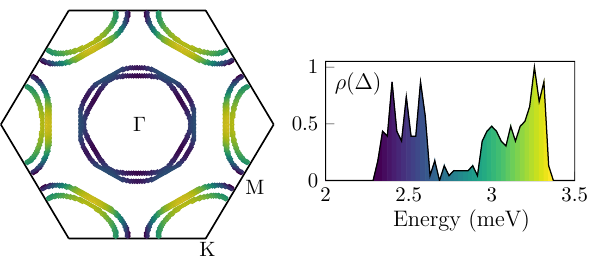}
                \caption{The superconducting gap on the Fermi surface $\Delta_{n\mathbf{k}}$ of bilayer NbSe$_2$, obtained from anisotropic Eliashberg calculations, at 3 K. $\rho(\Delta)$ is the normalized distribution of the gap in energy space, also serving as a color bar legend for $\Delta_{n\mathbf{k}}$.}
                \label{fig:NbSe2_gap}
\end{figure}

\begin{figure*}[t]
                \includegraphics[width=0.9\linewidth]{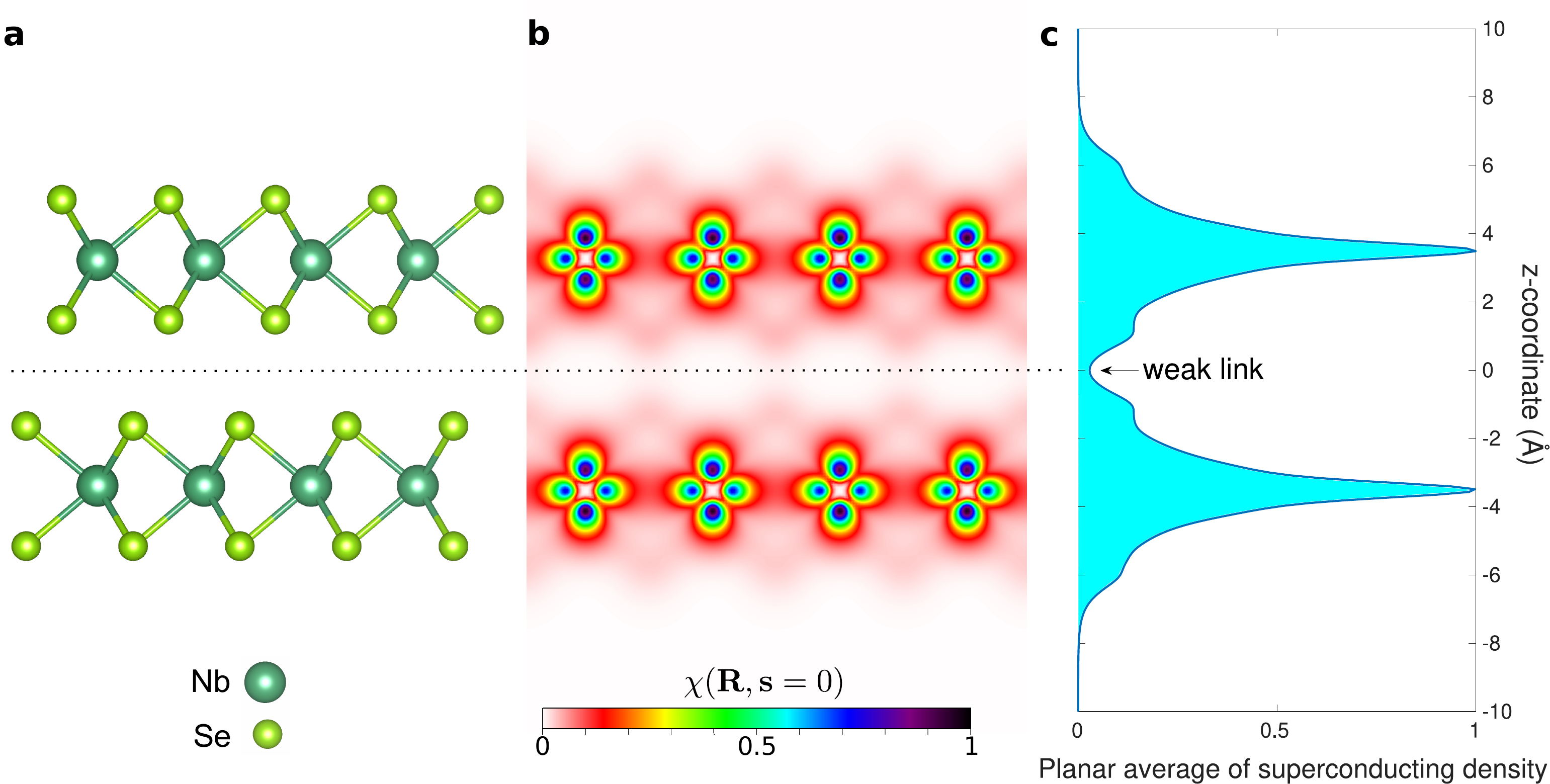}
                \caption{Prototype atomically-thin Josephson junction based on bilayer NbSe$_2$. (a) Crystal structure of bilayer NbSe$_2$, (b) normalized real-space superconducting density $\chi(\textbf{R},\mathbf{s}=0)$, and (c) evolution of the planar average of $\chi$ along the out-of-plane direction, revealing a weak link in the vdW gap between the two superconducting sheets, enabling the Josephson effect.}
                \label{fig:NbSe2_JJ}
\end{figure*}

Having established a proof-of-concept of the real-space approach on monolayer materials, we now turn to a bilayer (BL) structure, in order to quantify the effect of interlayer hybridization on the order parameter. As the test system, we have selected the transition metal dichalcogenide (TMD) material niobium diselenide (NbSe$_2$) in its 2H structural phase. As NbSe$_2$ lacks an inversion center in monolayer form (i.e., 1H layer), spin-orbit coupling (SOC) leads to Ising-type superconducting pairing, with a strong enhancement of the upper critical magnetic field \cite{Xi2016}. However, in BL form, the presence of an inversion center (cf.\ the structure shown in Figure \ref{fig:NbSe2_JJ}a) significantly reduces the effect of SOC on the superconducting properties.

The superconducting $T_c$ of NbSe$_2$ generally decreases in the ultrathin limit, dropping from the bulk value of 7 K to 5 K in the bilayer form and further down to 3 K in the monolayer \cite{Xi2016,khestanova2018}. Our first-principles calculations yielded a total electron-phonon coupling constant of $\lambda=1.05$, with a corresponding McMillan-Allen-Dynes $T_c$ of 14.6 K. The potential occurrence of ferromagnetic spin fluctuations in NbSe$_2$ has recently been proposed to explain the discrepancy between the \textit{ab initio} and experimental $T_c$ values \cite{Das2023}. 

As the next step in our investigation we evaluated the superconducting properties of BL NbSe$_2$ by solving the anisotropic Migdal-Eliashberg equations as implemented in the EPW code \cite{Giannozzi_2017,PhysRevB.88.085117,PONCE2016116} (computational details are provided in the SM). The resulting gap function $\Delta_{n\mathbf{k}}$ of BL NbSe$_2$, shown in Figure \ref{fig:NbSe2_gap}, displays significant anisotropy, with a weaker gap ($\sim 2.5$ meV) around the $\Gamma$ point, and a stronger one ($\sim 3.2$ meV) around the K point. This anisotropy of the gap is also clearly manifested in the distribution $\rho(\Delta)$ shown in Figure \ref{fig:NbSe2_gap}. A similar anisotropic gap distribution has been reported in previous studies on monolayer NbSe$_2$ \cite{PhysRevB.99.161119,Das2023}. It should be noted that including spin fluctuations in NbSe$_2$ mainly leads to a uniform shift in gap energies, leaving the distribution of the gap on the Fermi surface largely unchanged \cite{Das2023}.

The real-space superconducting density $\chi$, calculated from this gap distribution, is presented in Figure \ref{fig:NbSe2_JJ}b. Notably, it prominently reflects the shape of the Nb-4$d_{z^2}$ orbitals, which dominate the electronic states near the Fermi level of NbSe$_2$, further strengthening the connection between real-space superconducting properties and bonding characteristics. 

Generally, $\chi$ in this system reaches its maximum around the Nb planes. To further characterize this behavior, we have computed the planar average of the density along the out-of-plane direction, as shown in Figure \ref{fig:NbSe2_JJ}c. The results reveal a pronounced central peak at the $z$-coordinate of the Nb atoms. Additionally, there is a noticeable asymmetry between the density profile around the vdW gap and at the interfaces of the layers with the vacuum, highlighting the role of interlayer interactions. 

An essential feature of this system, unveiled through our real-space approach, is the emergence of superconducting tunneling across the vdW gap. The calculated density profile reveals that the vdW gap functions as an intrinsic weak link, enabling Josephson tunneling of Cooper pairs between the two layers. This result provides the first direct theoretical evidence of a weak link in the vdW gap of BL NbSe$_2$, confirming its potential to function as an atomically thin Josephson junction.

This result obtained for BL NbSe$_2$ demonstrates that Josephson tunneling requires sufficient overlap between the normal-state electronic wave functions within the tunneling region. In multilayer structures of 2D materials, this interlayer hybridization can be controlled either by selecting an appropriate barrier material, e.g.\ graphene \cite{Kim2017}, or using the vdW gap itself \cite{Yabuki2016,Li2020,Farrar2021}. The latter approach, explored in this work, represents the most fundamental barrier configuration, allowing for precise regulation of interlayer hybridization through tunable system parameters such as interlayer spacing, stacking order, and twist angle. Barrier engineering through twisting has been experimentally realized in several layered TMD superconductors, including NbSe$_2$, confirming that the twist angle modulates the critical Josephson current of the junction \cite{Li2020,Farrar2021}. Furthermore, recent experiments on twisted heterostructures, such as monolayer NbSe$_2$/graphene, have revealed that the superconducting gap can be selectively and anisotropically modulated by varying the twist angle \cite{Naritsuka2025}. 

Therefore, our results on BL NbSe$_2$ provide a theoretical foundation for these experiments and offer a framework for designing ultrathin Josephson junctions. This has significant implications for emerging quantum technologies, as Josephson junctions are fundamental building blocks of advanced quantum devices, particularly superconducting qubits \cite{Martinis2004,Arute2019,science1084,PhysRevLett.127.180501,doi:10.1146/annurev-conmatphys-031119-050605}. Current state-of-the-art qubit technologies rely on superconducting junctions fabricated from bulk materials, which can introduce disorder and defects \cite{Liu2019}. Scattering from these defects typically disrupts Cooper pairs, generating dissipative quasiparticles that induce fluctuations in qubit frequencies and degrade quantum coherence. In contrast, Josephson junctions based on superconducting 2D materials, such as BL NbSe$_2$, offer exceptionally clean interfaces with atomic precision, minimizing disorder and enhancing qubit coherence \cite{Liu2019}. Furthermore, the ability to actively modulate junction materials using techniques such as strain \cite{Chen2020,doi:10.1021/acs.nanolett.4c02421} and gating \cite{deVries2021,Rodan-Legrain2021} offers a powerful approach for implementing quantum operations.

In conclusion, we have presented a computational framework based on \textit{ab initio} Migdal-Eliashberg theory to efficiently evaluate real-space superconducting properties by combining the superconducting gap spectrum with the Bloch wave functions of the Fermi states. All the physical quantities required for the calculation of the real-space superconducting gap distribution are readily implemented in the Quantum ESPRESSO and EPW packages, making this approach broadly accessible to the community.

We have applied this framework to a selection of atomically thin materials to demonstrate the advantages of a real-space perspective in understanding superconductivity at the nanoscale. Our comparative analysis of two gallenene polymorphs has revealed distinct Cooper-pair distributions, with the out-of-plane component defining superconducting layer thicknesses that capture the evanescence of the superconducting state beyond the atomic layer. 
 
For bilayer NbSe$_2$, we revealed the vdW gap as an intrinsic weak link in the context of Cooper-pair tunneling, underscoring its potential as an atomically thin Josephson junction. While previous experimental studies have demonstrated Josephson coupling in thicker flakes of NbSe$_2$ \cite{Yabuki2016,Kim2017,Li2020,Farrar2021}, our work shows that such coupling persists even in the bilayer limit. This highlights the robustness of interlayer Cooper-pair tunneling in the most minimal vdW Josephson junction architectures. Such junctions hold significant promise for emerging quantum technologies, particularly to enhance the coherence of superconducting qubits by minimizing disorder and enabling active tuning \cite{Liu2019}.

Beyond NbSe$_2$ homobilayers, several other TMD-based systems are compelling candidates to apply our real-space approach to, due to their pronounced spatial anisotropy and strong interlayer interactions. TaS$_2$-based natural vdW heterostructures, such as the 6R-phase, offer a promising platform for vdW-type junctions, where insulating 1T layers serve as barriers between superconducting 1H planes \cite{Achari2022,PhysRevLett.133.056001}. Additionally, recent studies on twisted W-based TMDs, including WSe$_2$ \cite{Guo2025,Xia2025} and WTe$_2$ \cite{Wang2022,Yu2023}, have uncovered unconventional superconductivity shaped by strong electronic correlations, leading to pronounced spatial features such as one-dimensional Luttinger liquid behavior. Another notable example is proximity-induced superconductivity at the surface of a topological insulator interfaced with NbSe$_2$ \cite{PhysRevB.109.L140503}. These systems highlight scenarios where real-space anisotropy and Cooper-pair tunneling across interfaces are essential for understanding and tailoring superconducting properties.

The real-space approach presented here therefore offers distinct advantages in visualizing the spatial distribution of the superconducting state and understanding the impact of various factors such as anisotropy, interlayer hybridization, and local features. This detailed spatial information is crucial for the rational design and optimization of advanced superconducting heterostructures and quantum devices.

\begin{acknowledgments}
\noindent J.B. acknowledges support as a Senior Postdoctoral Fellow of Research Foundation-Flanders (FWO) under Fellowship No.\ 12ZZ323N. The computational resources and services were provided by the VSC (Flemish Supercomputer Center), funded by the FWO and the Flemish Government -- department EWI. 
\end{acknowledgments}

\bibliography{Refs}

\end{document}